\documentclass[twocolumn,floatfix,eqsecnum,aps]{revtex4-2}
\usepackage{graphicx}
\usepackage{amsmath}
\usepackage{amssymb}
\usepackage{bm}
\usepackage{hyperref}

\usepackage{color}

\begin{document}

\title{Poor-man's Majorana edge mode enabled by specular Andreev reflection}
\author{C. W. J. Beenakker}
\affiliation{Instituut-Lorentz, Universiteit Leiden, P.O. Box 9506, 2300 RA Leiden, The Netherlands}
\date{September 2024}

\begin{abstract}
It is known that the surface of a three-dimensional topological insulator (3D TI) supports a chiral Majorana edge mode at the interface between a superconductor and a magnetic insulator. The complexity of the materials combination is such that this state has not yet been observed.  Here we show that a helical Majorana edge mode appears even in the absence of the magnetic insulator, if the Fermi level of the massless surface electrons is at the Dirac point. Specular Andreev reflection of Dirac fermions is at the origin of the effect. The simplified geometry may favor experimental observation of the helical Majorana mode, although it lacks the topological protection of its chiral counterpart. 
\end{abstract}
\maketitle

\section{Introduction}

\begin{figure}[tb]
\centerline{\includegraphics[width=1\linewidth]{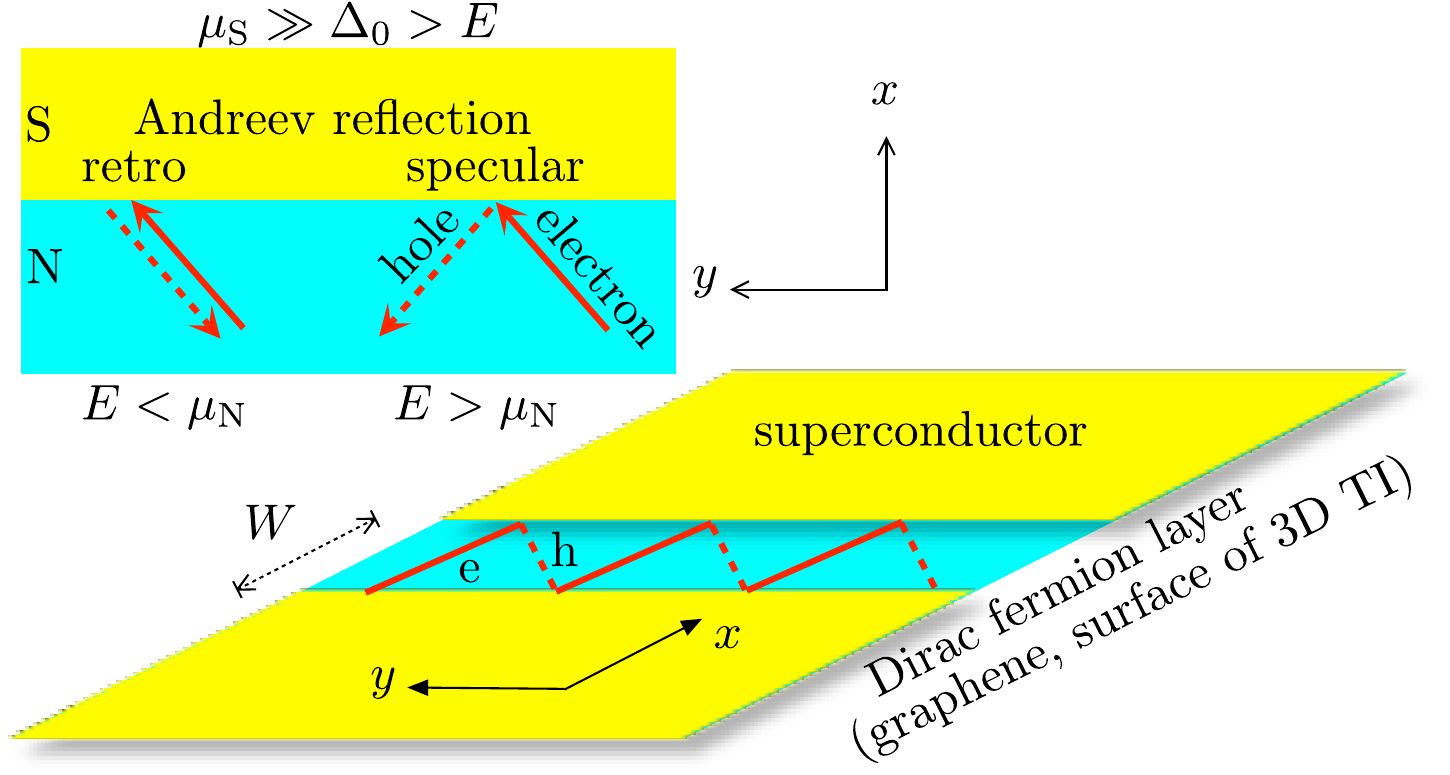}}
\caption{Upper panel: Retro-reflection versus specular reflection of a Dirac fermion at a normal--superconductor (NS) interface.
%(superconducting gap $\Delta_0$ greater than excitation energy $E$, chemical potential $\mu_{\rm S}$ in the superconductor much greater than $\Delta_0$). 
Lower panel: If the chemical potential $\mu_{\rm N}$ in the normal region is less than the excitation energy $E$, specular Andreev reflection allows for a Dirac fermion to propagate between two NS interfaces, in an equal-weight electron-hole superposition. Here we show that such a charge-neutral mode can exist as well in the case of a \textit{single} NS interface. 
The edge state is bound to the NS interface because it lies outside of the Dirac cone of bulk states.
}
\label{fig_layout}
\end{figure}

Andreev reflection \cite{And64} is the process by which an electron incident on a superconductor is reflected as a hole --- the missing charge of $2e$ accounted for by a Cooper pair in the superconducting condensate. Because the hole retraces the path of the electron, one speaks of \textit{retro}-reflection (all velocity components change sign). This applies to massive electrons, governed by the Schr\"{o}dinger equation. 

The Dirac equation of massless quasiparticles also allows for Andreev reflection by a superconductor, with a qualitative difference: At energies near the Dirac point only the velocity component perpendicular to the interface changes sign, the parallel component is unchanged so that the reflection is \textit{specular} \cite{Bee06,Efe16,Wan21}. As illustrated in Fig.\ \ref{fig_layout}, specular Andreev reflection enables a new transport mode in a narrow channel, charge-neutral because it is an equal-weight electron-hole superposition \cite{Tit07,Gre07}.

From this description the presence of a pair of opposite normal-superconductor (NS) interfaces seems essential, to allow for the repeated specular Andreev reflections that are needed for a charge-neutral mode. Here we will show that a single NS interface suffices. When the Fermi level lines up with the Dirac point, the spectrum consists of a Dirac cone with a helical edge mode outside of the cone. The charge expectation value of the edge mode vanishes for all energies $E$ up to the gap $\Delta_0$. It has the real wave function of a Majorana fermion \cite{Wil09,Bee16}. 

Because the helical edge mode coexists with bulk modes there is no topological protection, unlike the chiral Majorana edge mode that is predicted to exist on the surface of a 3D TI, at the interface between a superconductor and a magnetic insulator \cite{Fu08,Has10,Qi11}. 

The two types of edge modes are intimately related: The chiral Majorana mode evolves \textit{continuously} into one branch of the helical mode as the magnetic gap is reduced below the superconducting gap. That the Majorana edge mode does not vanish by merging with the bulk states once the gap closes is a weak form of protection. Following a terminology \cite{Lei12} used for Majorana zero-modes we might speak of a ``poor man's'' Majorana edge mode. Since topologically protected Majorana edge modes are still elusive, the option to study some of their properties in a more readily accessible system is of interest.

\section{Edge mode at a single NS interface}

\subsection{Model Hamiltonian}

Non-degenerate massless Dirac fermions exist on the two-dimensional (2D) surface of a 3D topological insulator (3D TI) \cite{Has10,Qi11}. We will focus on that system in what follows, but much of the physics applies also to graphene --- if we may neglect intervalley scattering in the carbon monolayer. The low-energy dynamics on the surface (in the $x$--$y$ plane) is governed by the 2D Dirac Hamiltonian,
\begin{equation}
H_{\rm Dirac}=v_{\rm F}(p_x\sigma_x+p_y\sigma_y),
\end{equation}
with $v_{\rm F}$ the Fermi velocity, $\bm{p}=-i\hbar\partial/\partial\bm{r}$ the momentum operator, and $\sigma_\alpha$ a Pauli spin matrix.

A spin-singlet, \textit{s}-wave superconductor deposited on the 2D surface induces a pair potential $\Delta$, which gaps the surface states. The pair potential enters as a term that mixes electrons and holes in the Bogoliubov-De Gennes Hamiltonian,
\begin{equation}
\begin{split}
&H_{\rm BdG}=H_0+\Delta(x)\sigma_0\tau_x,\\
&H_{0}=v_{\rm F}(p_x\sigma_x+p_y\sigma_y)\tau_z-\mu(x)\sigma_0\tau_z.
\end{split}
\label{HBdG}
\end{equation}
The $\sigma$ and $\tau$ Pauli matrices act, respectively, on the spin and electron-hole degree of freedom (with $\sigma_0$ and $\tau_0$ the  $2\times 2$ unit matrix). We set $\Delta(x)=\Delta_0>0$ for $x>0$ (region S, with $\xi_0=\hbar v_{\rm F}/\Delta_0$ the superconducting coherence length) and $\Delta(x)=0$ for $x<0$ (region N). The chemical potential $\mu(x)$ in the two regions may be different, equal to $\mu_{\rm S}$ and $\mu_{\rm N}$, respectively. We will set $v_{\rm F}$ and $\hbar$ to unity in most equations.

Because of translational invariance along the NS interface, in the $y$-direction, the parallel momentum component $p_y=\hbar k_y\equiv q$ is a good quantum number. We seek an eigenstate $\psi(x)e^{iqy}$ of $H_{\rm BdG}$ at energy $|E|<\Delta_0$ that decays exponentially away from the interface, for $|x|\rightarrow\infty$.

\subsection{Dispersion relation}

The eigenvalue problem for a given parallel momentum $q$ is
\begin{align}
&-i\sigma_x\tau_z\psi'(x)=\\
&=\begin{cases}
[E\sigma_0\tau_0+(\mu_{\rm N}\sigma_0-q\sigma_y)\tau_z]\psi(x),&x<0,\\
[E\sigma_0\tau_0+(\mu_{\rm S}\sigma_0-q\sigma_y)\tau_z-\Delta_0\sigma_0\tau_x]\psi(x),&x>0.
\end{cases}\nonumber
\end{align}
The solution is
\begin{equation}
\psi(x)=\begin{cases}
e^{\Xi_{\rm N} x}\psi(0),&x<0,\\
e^{\Xi_{\rm S}x}\psi(0),&x>0,
\end{cases}
\end{equation}
with matrices
\begin{equation}
\begin{split}
&\Xi_{\rm N}=iE\sigma_x\tau_z+(q\sigma_z+i\mu_{\rm N}\sigma_x)\tau_0,\\
&\Xi_{\rm S}=iE\sigma_x\tau_z+(q\sigma_z+i\mu_{\rm S}\sigma_x)\tau_0+\Delta_0\sigma_x\tau_y.
\end{split}
\end{equation}

Denote the \textit{right} eigenvectors of $\Xi_{\rm S}$ with eigenvalue having a negative real part (so producing a decaying state for $x\rightarrow\infty$) by $u_1$ and $u_2$. The state $\psi(0)$ must be a superposition of $u_1$ and $u_2$ which also decays for $x\rightarrow-\infty$, so it must be orthogonal to the \textit{left} eigenvectors $ v_1, v_2$ of $\Xi_{\rm N}$ with eigenvalue having a negative real part. This requires that the $2\times 2$ matrix $M$ with elements $M_{ij}=\langle v_i|u_j\rangle$ has determinant equal to zero. 

\begin{figure}[tb]
\centerline{\includegraphics[width=0.7\linewidth]{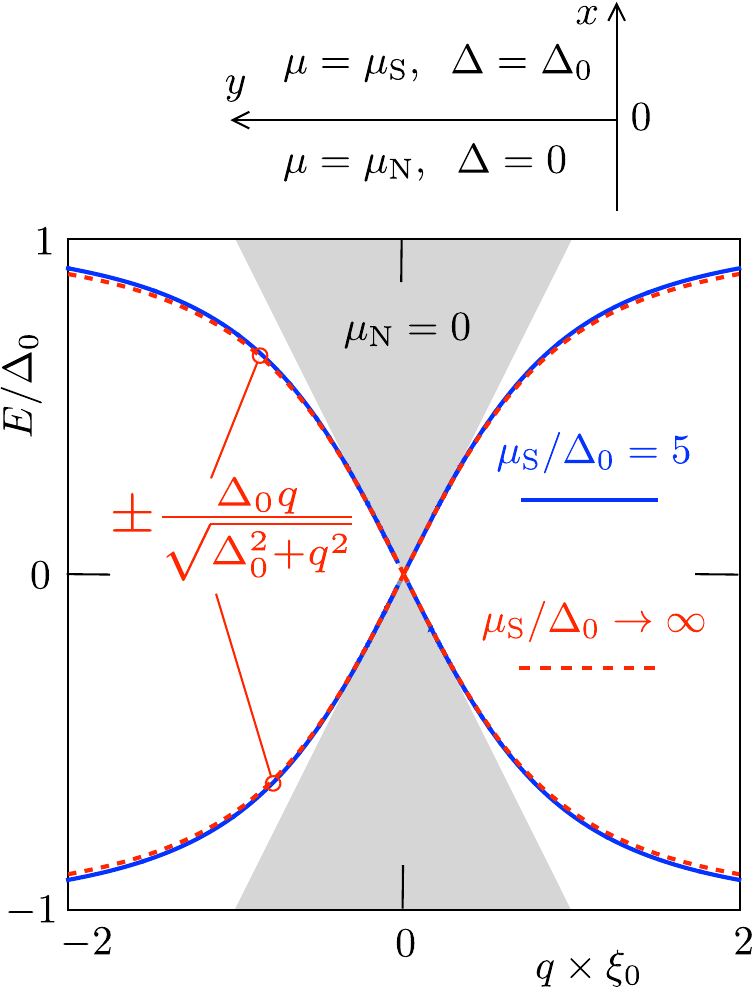}}
\caption{Blue curves: Dispersion relation of the edge mode at a single NS interface (geometry shown at the top, $q$ is the parallel momentum component). This is the numerical result for $\mu_{\rm S}=5\Delta_0$ and $\mu_{\rm N}=0$. The analytic limit \eqref{Epmresult} for $\mu_{\rm S}/\Delta_0\rightarrow\infty$ is the red dashed curve. The Dirac cone of bulk states is indicated in grey. 
}
\label{fig_spectrum1}
\end{figure}

\begin{figure}[tb]
\centerline{\includegraphics[width=0.7\linewidth]{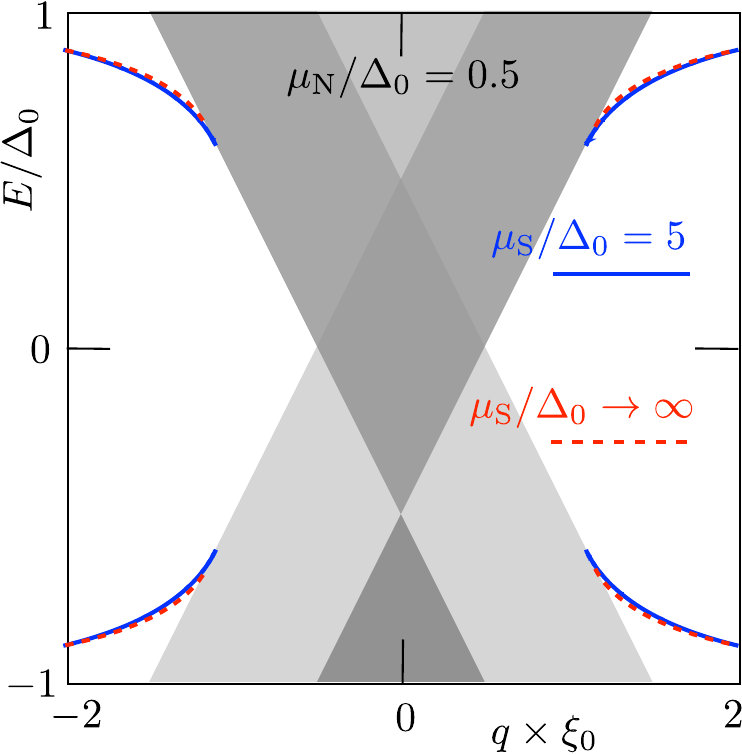}}
\caption{Same as Fig.\ \ref{fig_spectrum1}, for $\mu_{\rm N}=0.5\,\Delta_0$. The electron and hole Dirac cones (two shades of grey) have a relative displacement of $2\mu_{\rm N}$. 
}
\label{fig_spectrum2}
\end{figure}

Solving $\det M=0$ for $E$ at a given $q$ gives the edge mode dispersion relation. The general expression is complicated, we solve it numerically. In Figs.\ \ref{fig_spectrum1} and \ref{fig_spectrum2} we plot the resulting dispersion for $\mu_{\rm S}=5\Delta_0$ and two values of $\mu_{\rm N}$. In each case the edge modes lie outside of the Dirac cone of bulk states, to which they connect tangentially. (This is a general requirement for edge-bulk reconnections \cite{Hal14,Bee24}.) The edge mode exists over the entire range of parallel momenta for $\mu_{\rm N}=0$, but for nonzero $\mu_{\rm N}$ it only exists for sufficienty large $|q|$.

\subsection{Limit of large chemical potential in S}

The superconductor which induces a pair potential on the surface of the 3D TI will also raise the local chemical potential, to a value $\mu_{\rm S}$ large compared to $\Delta_0$. It is therefore natural to take the limit $\mu_{\rm S}/\Delta_0\rightarrow\infty$, when the dispersion relation can be computed in closed form.

In the large-$\mu_{\rm S}$ limit the eigenvectors $u_1,u_2,v_1,v_2$ introduced in the previous subsection reduce to
\begin{subequations}
\label{uveqs}
\begin{align}
&u_1=\bigl(E+i\sqrt{\Delta_0^2-E^2},E+i\sqrt{\Delta_0^2-E^2},\Delta_0,\Delta_0\bigr),\\
&u_2=\bigl(E-i\sqrt{\Delta_0^2-E^2},-E+i\sqrt{\Delta_0^2-E^2},\Delta_0,-\Delta_0\bigr),\\
&v_1=\bigl(\sqrt{q^2-(\mu_{\rm N}+E)^2}-q,i(\mu_{\rm N}+E),0,0\bigr),\\
&v_2=\bigl(0,0,\sqrt{q^2-(\mu_{\rm N}-E)^2}-q,i(\mu_{\rm N}-E)\bigr).
\end{align}
\end{subequations}
The eigenvectors $u_1,u_2$ are $q$-independent, because the term $q\sigma_z\tau_0$ in $\Xi_{\rm S}$ anticommutes with the term $i\mu_{\rm S}\sigma_x\tau_0$ and does not contribute in the large-$\mu_{\rm S}$ limit.

The solution of $\det M=0$, $M_{ij}=\langle v_i|u_j\rangle$, gives the dispersion relation
\begin{align}
E_\pm(q)={}&\frac{\pm\Delta_0}{\sqrt{2}\sqrt{\Delta_0^2+q^2}}\nonumber\\
&\times\sqrt{q^2+\mu_{\rm N}^2+\sqrt{(q^2-\mu_{\rm N}^2)^2-4\Delta_0^2\mu_{\rm N}^2}},\label{Epmqresult}
\end{align}
in the interval $|q|>q_c$, where $q_c>0$ is the solution of
\begin{equation}
E(q_c)=q_c-\mu_{\rm N}\Rightarrow q_c\approx\Delta_0(\mu_{\rm N}/\Delta_0)^{1/3}+\tfrac{2}{3}\mu_{\rm N}+\cdots
\end{equation}

The large-$\mu_{\rm S}$ asymptotics \eqref{Epmqresult} is plotted in Figs.\ \ref{fig_spectrum1} and \ref{fig_spectrum2} (red dashed curves). It is close to the finite-$\mu_{\rm S}$ results (blue solid curves) for $\mu_{\rm S}\gtrsim 5\Delta_0$.

\subsection{Fermi level aligned with the Dirac point}

A simple expression for the edge state results when $\mu_{\rm N}=0$ (still taking $\mu_{\rm S}/\Delta_0\rightarrow\infty$). The branch $\psi_+$ with $dE/dq>0$ (normalized to unity) has the form \cite{note}
\begin{subequations}
\label{state_result}
\begin{align}
&\psi_+(x)=C\begin{pmatrix}
i\cos(\alpha/2))\\
-\sin(\alpha/2)\\
i\cos(\alpha/2)\\
\sin(\alpha/2)
\end{pmatrix}\times\begin{cases}
e^{q x\cos\alpha}&x<0,\\
e^{-\Delta_0 x\sin\alpha}&x>0,
\end{cases}\nonumber\\
&C=\sqrt{q\cos\alpha+\Delta_0\sin\alpha},\\
&\alpha=\arccos\left(\frac{q}{\sqrt{\Delta_0^2+q^2}}\right)\in(0,\pi),
\end{align}
\end{subequations}
at energy
\begin{equation}
E_+=\Delta_0\cos\alpha=\frac{\Delta_0 q}{\sqrt{\Delta_0^2+q^2}}.\label{Epmresult}
\end{equation}
The state decays into N and S with decay lengths $\lambda_{\rm N}=(q\cos\alpha)^{-1}$ and $\lambda_{\rm S}=\xi_0/\sin\alpha$, respectively. The branch $\psi_-$ with $dE/dq<0$ follows from chiral symmetry,
\begin{equation}
\psi_-(x)=\sigma_z\tau_z\psi_+(x),\;\;E_-=-\Delta_0\cos\alpha.
\end{equation}

The state $\psi_+={{\psi_{e}}\choose{\psi_{h}}}$ represents a charge-neutral helical mode, equal weight of electron and hole components $\psi_e,\psi_h$ for all energies up to the superconducting gap. The real representation characteristic of  Majorana fermion mode results upon a unitary transformation,
\begin{align}
\psi_+\mapsto{}& \frac{1}{\sqrt 2}\begin{pmatrix}
1&-1\\
i&i
\end{pmatrix}\begin{pmatrix}
\psi_e\\
\psi_h
\end{pmatrix}\label{EMajorana}\\
={}&-C\sqrt 2\begin{pmatrix}
0\\
\sin(\alpha/2)\\
\cos(\alpha/2)\\
0
\end{pmatrix}\times\begin{cases}
e^{q x\cos\alpha}&x<0,\\
e^{-\Delta_0 x\sin\alpha}&x>0.
\end{cases}\nonumber
\end{align}

%For nonzero $\mu_{\rm N}$ the charge expectation value is computed from
%\begin{equation}
%\langle Q\rangle=\langle\psi|e\sigma_0\tau_z|\psi\rangle=-e\frac{dE}{d\mu_{\rm N}}.
%\end{equation}
%It equals a full $\pm e$ when the edge mode detaches from the bulk band in N (at $q=q_c$), and then decays to zero as the mode approaches the gap in S.

\section{Conversion of chiral to helical Majorana  mode}

Addition of the magnetization $M(x)\sigma_z\tau_0$ to the Hamiltonian $H_0$ in Eq.\ \eqref{HBdG} gaps the normal region, if $M=M_0$ for $x<0$, while $M=0$ for $x>0$. This describes the interface between a magnetic insulator and a superconductor, both deposited on the surface of a 3D TI. A chiral Majorana edge mode exists at the interface, with propagation direction set by the sign of $M_0$ \cite{Fu08,Has10,Qi11}. How is the chiral edge mode related to the helical edge mode without the magnetic insulator?

A nonzero $M_0$ modifies the eigenvectors $v_1,v_2$ in Eq.\ \eqref{uveqs}, which are now given by
\begin{subequations}
\label{veqs}
\begin{align}
&v_1=\bigl(\sqrt{q^2+M_0^2-(\mu_{\rm N}+E)^2}-q,i(\mu_{\rm N}+E-M_0),0,0\bigr),\\
&v_2=\bigl(0,0,\sqrt{q^2+M_0^2-(\mu_{\rm N}-E)^2}-q,i(\mu_{\rm N}-E+M_0)\bigr).
\end{align}
\end{subequations}

\begin{figure}[tb]
\centerline{\includegraphics[width=0.7\linewidth]{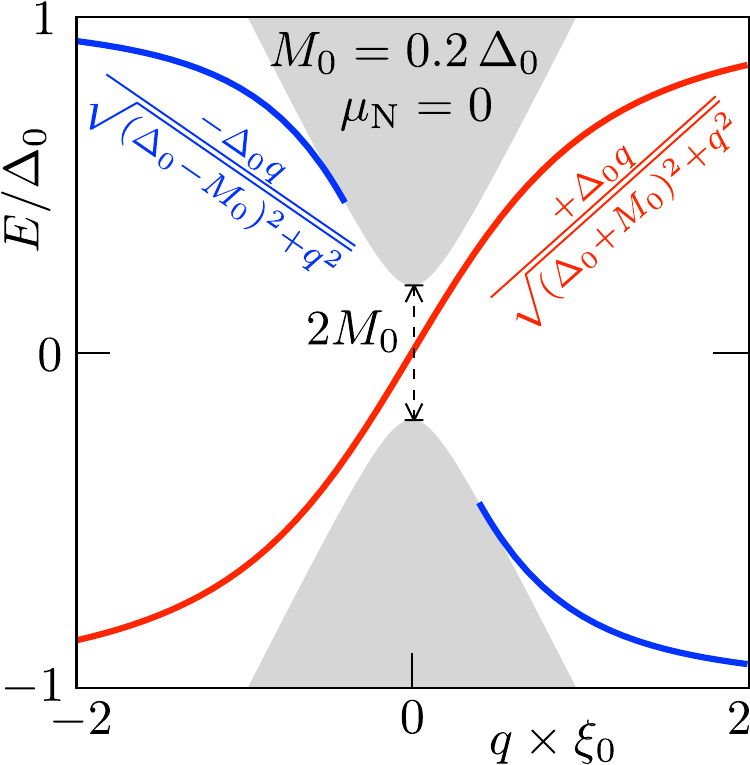}}
\caption{Edge mode spectrum for the case that the normal region is gapped by a magnetic insulator (computed from Eq.\ \eqref{Epmqgap} for magnetization $M_0=0.2\,\Delta_0$, with $\mu_{\rm N}=0$, $\mu_{\rm S}\rightarrow\infty$). The chiral Majorana mode (red curve) extends through the gap and will evolve into one branch of the helical Majorana mode when $M_0\rightarrow 0$. The counterpropagating branch (blue curve) does not connect through the gap, it tangentially connects to the bulk spectrum (shaded grey, $E^2>q^2+M_0^2$) at nonzero energy.
}
\label{fig_spectrum3}
\end{figure}

A closed-form expression for the spectrum results in the large-$\mu_{\rm S}$ limit with $\mu_{\rm N}=0$. Taking $0<M_0<\Delta_0$ we find that the spectrum consists of the counterpropagating branches
\begin{subequations}
\begin{equation}
E_\pm(q)=\frac{\pm\Delta_0 q}{\sqrt{(\Delta_0\pm M_0)^2+q^2}},
\end{equation}
where $E_+(q)$ exists for all $q$, while $E_-(q)$ only exists for 
\begin{equation}
E_-^2>M_0\Delta_0\Rightarrow q^2>M_0(\Delta_0-M_0).
\end{equation}
\label{Epmqgap}
\end{subequations}

Fig.\ \ref{fig_spectrum3} shows this spectrum for $M_0/\Delta_0=0.2$. For $M_0>\Delta_0$ solely the chiral edge mode is present, once $M_0$ drops below $\Delta_0$ a counterpropagating mode appears. The chiral mode extends through the gap for any $M_0$, the counterpropagating mode only in the limit $M_0\rightarrow 0$, when the helical pair of Majorana modes is formed.

\section{Conclusion}

\begin{figure}[tb]
\centerline{\includegraphics[width=1\linewidth]{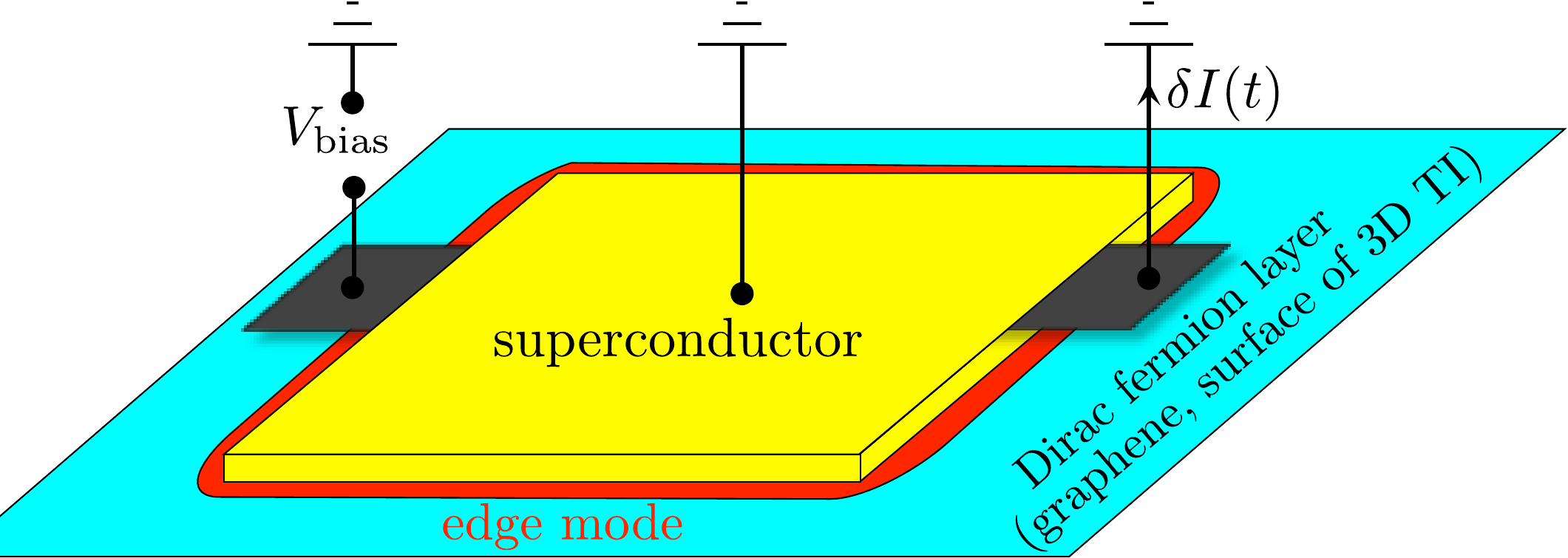}}
\caption{Geometry to detect the edge mode encircling a voltage biased and grounded superconductor via the current fluctuations (shot noise) measured at a remote contact.
}
\label{fig_experiment}
\end{figure}

We have discovered that a superconductor on a massless Dirac fermion layer (graphene or 3D TI surface) binds a helical edge mode, propagating as a coherent electron-hole superposition via repeated specular Andreev reflections at the normal-superconductor interface. The mode decays into the superconductor because of the gap $\Delta_0$ and it decays into the normal region because it lies outside of the Dirac cone. Unlike the known chiral Andreev or Majorana edge modes \cite{Ero05,Par14,Zha20,Qi10,Li20}, here no magnetic field or magnetic insulator is needed to gap the Dirac fermions: the helical edge mode coexists with a gapless normal region.

When the Fermi level in the normal region lines up with the Dirac point, the edge mode becomes a charge-neutral Majorana mode, described by the real wave function \eqref{EMajorana}. Although it carries no electrical current on average, it can be detected via the current fluctuations. 

A geometry for such a shot noise measurement is shown in Fig.\ \ref{fig_experiment}. An unpaired Majorana mode produces a shot noise power of $\tfrac{1}{2}e^2/h$ per $eV$ of voltage bias \cite{Akh11}. This was studied in Ref.\ \onlinecite{Gne15} in the case that the Majorana mode is topologically protected, at the edge of a chiral \textit{p}-wave superconductor. The quantized shot noise then persists for voltages up to $\Delta_0$, without any sensitivity to decoherence or impurity scattering.

Here one would also want to apply voltages comparable to $\Delta_0$, to ensure that the edge mode is tightly bound to the superconductor [the penetration depth in the normal region at energy $E$ is of order $\xi_0(\Delta_0/E)^2$]. Because the edge mode coexists with gapless Dirac fermions, it lacks the topological protection of a gapped system. One might surmise that long-range scattering will not be effective at coupling the large-$q$ edge modes to the small-$q$ bulk modes. In any case the contact separation in Fig.\ \ref{fig_experiment} would have to smaller than this coupling length. While such a device would not be helpful for quantum applications, it could be a way to study a charge-neutral edge mode in a ``poor man's'' system --- as simple as proximitized graphene.

\acknowledgments

This project has received funding from the European Research Council (ERC) under the European Union's Horizon 2020 research and innovation programme. Discussions with I. Adagideli, A. R. Akhmerov, and F. Hass\-ler are gratefully acknowledged.


\begin{thebibliography}{99}
\bibitem{note} As far as I know, the edge state \eqref{state_result} was not noticed in the large number of papers that have studied the NS interface in graphene or topological insulators. I was suspicious that such a simple result could have been missed, and worried that it might be mistaken. But there is a simple direct check for its validity: For $x<0$, one readily checks that $H_{\rm Dirac}\psi_e=E\psi_e$, $H_{\rm Dirac}\psi_h=-E\psi_h$, while at $x=0$ the electron and hole wave functions are related by the boundary condition $\psi_h=\exp[-i\sigma_x\arccos(E/\Delta_0)]\psi_e$.
\end{thebibliography}
\end{document}